\documentclass[prb,aps,letter,superscriptaddress,showpacs,twocolumn,floatfix]{revtex4-1}

\usepackage{graphicx}
\usepackage[ansinew]{inputenc}
\usepackage{array}
\usepackage{color}
\usepackage{amsmath}
\usepackage{amsxtra}
\usepackage{amstext}
\usepackage{amssymb}
\usepackage{latexsym}
\begin{document}

\title{Imaging charge density fluctuations in graphene using Coulomb blockade spectroscopy}

\author{A.	Deshpande}
\altaffiliation[Current Address: ]{Department of Materials Science and Engineering, Northwestern University, Evanston, IL, 60208 USA.}
\affiliation{Department of Physics, University of Arizona, Tucson, AZ, 85721 USA.}
\author{W. Bao}
\author{H. Zhang}
\author{Z. Zhao}
\author{C. N. Lau}
\affiliation{Department of Physics, University of California at Riverside, Riverside, CA 92521 USA.}
\author{B. J. LeRoy}
\email{leroy@physics.arizona.edu}
\affiliation{Department of Physics, University of Arizona, Tucson, AZ, 85721 USA.}

\date{\today}

\begin{abstract}
Using scanning tunneling microscopy, we have imaged local charge density fluctuations in monolayer graphene.  By placing a small gold nanoparticle on the end of the STM tip, a charge sensor is created.  By raster scanning the tip over the surface and using Coulomb blockade spectroscopy, we map the local charge on the graphene.  We observe a series of electron and hole doped puddles with a characteristic length scale of about 20 nm.  Theoretical calculations for the correlation length of the puddles based on the number of impurities are in agreement with our measurements.
\end{abstract}


\maketitle
Graphene is the two dimensional form of carbon consisting of a single layer of carbon atoms arranged in a hexagonal lattice. This simple geometry is a repository of exceptional electronic, mechanical, and chemical properties. These properties emerge from the linear band structure of graphene. The charge carriers in graphene behave as massless Dirac fermions and exhibit very high mobility \cite{novoselov2005,neto2009}. This makes graphene very promising for device applications \cite{geim2009}. Efforts in this direction have yielded spectacular results from using graphene as a nanopore template for DNA \cite{dekker2010} to producing 30-inch monolayer graphene films on a copper substrate, detaching them and fabricating transparent electrodes \cite{bae2010}. However, one of the challenges in order to make graphene devices is controlling the electrostatic environment.  In particular, it has been shown that graphene on silicon dioxide (SiO$_2$) tends to form a series of electron and hole puddles near the Dirac point \cite{martin,deshpande2009,zhang2009}.  The puddles arise due to the random charged impurities on the graphene layer and in between the graphene and SiO$_2$. This reduces the mobility of graphene. The puddles are also partly responsible for the minimum conductivity in graphene. This limits the on-off ratio for graphene transistors as the `off' state is not reached \cite{novoselov2004} . Hence a complete understanding of puddles is vital for progress in graphene applications.

In this work, we use a nanoparticle on the end of a scanning tunneling microscope (STM) tip as a sensitive charge sensor to spatially map variations in the electron charge density. Our technique allows a direct quantitative measurement of the charge density fluctuations with a spatial resolution of several nanometers leading to an estimation of the number of impurities.  Previous studies of electron and hole puddles on graphene lacked the spatial and/or energy resolution to fully resolve the puddles and obtain information about their characteristic distribution. Charge density variations on a length scale of $\sim$150 nm have been measured with a single electron transistor (SET) \cite{martin}. The size of the SET and distance from the graphene limited the spatial resolution of these measurements.  Previous STM measurements of the electron and hole puddles lacked the energy resolution of our current technique because they relied on a shift in the local density of states\cite{deshpande2009,zhang2009}.  In these measurements, we use the sharp peaks in dI/dV measurements due to Coulomb blockade to significantly improve the energy resolution.

Graphene was first isolated by exfoliating it from 3D graphite and attaching it onto a SiO$_2$ substrate \cite{novoselov2004}. There are many alternative ways to make graphene, like epitaxy on silicon carbide substrates \cite{berger2004} or chemical vapor deposition growth on copper \cite{li2009} and nickel \cite{kim2009}.  However, the mechanical exfoliation technique still provides the highest quality devices.  Here, we have studied exfoliated graphene on a SiO$_2$ substrate using a STM operating in ultra-high vacuum at 4.6 K. Previous STM studies on exfoliated graphene have been instrumental in correlating its structure with the electronic properties \cite{zhang2008,deshpande2009,zhang2009}.  In this work, we report density fluctuations in graphene using Coulomb blockade spectroscopy measurements. Coulomb blockade is the suppression of electron tunneling when the charging energy $e^2/2C$, where $C$ is the capacitance of the nanoparticle, is unavailable to the electrons at low temperature and voltages, i.e. $k_BT, eV << e^2/2C$ \cite{devoret1990}.  With a nanoparticle at the end of the STM tip, the system exhibits the two junctions which are necessary to produce Coulomb blockade.  The tunneling gap between the nanoparticle and the surface acts as one junction, while the barrier between the tip and nanoparticle is the other. The high spatial resolution of the STM and the low temperature allows a spatially resolved measurement of Coulomb blockade.  As Coulomb blockade is sensitive to not only changes in capacitance but also the electrostatic environment of the nanoparticle, we are able to detect charge fluctuations on graphene using Coulomb blockade spectroscopy.

Graphene flakes were exfoliated from graphite on a Si substrate with 300 nm of thermally grown oxide. Ti/Au electrodes were then deposited on the graphene through a shadow mask \cite{Lau2010}.  This technique does not require the use of PMMA as a resist. Therefore it yields clean graphene devices that need no additional cleaning procedures prior to imaging with STM.  The graphene sample is transferred to the ultra-high vacuum (p $\leq 10^{-11}$ mbar) STM and cooled to 4.6 K. Electrochemically etched tungsten tips were used for imaging and spectroscopy.  A constant density of states for the tip was confirmed on a Au surface.  After confirmation of the density of states of the tip, a nanoparticle was attached to the end of it and spectroscopy measurements on graphene were performed.

\begin{figure}[t]
\includegraphics[width=0.45 \textwidth]{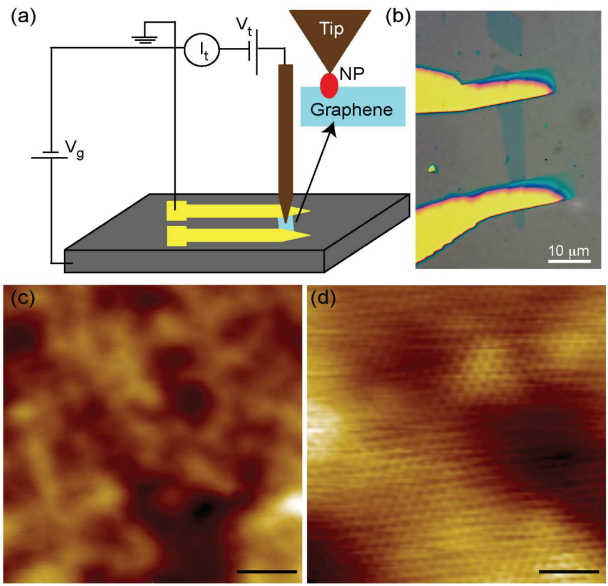}
\caption{\label{fig:schematic} (color online) (a) Schematic of the graphene sample with gold contacts on a  SiO$_2$ substrate. The nanoparticle at the tip apex is shown in the close up sketch. (b) Optical microscope image of the monolayer graphene flake and gold electrodes.  (c) Large scale STM image of the graphene surface. (0.5 V, 100 pA, 100 nm $\times$ 100 nm). The scale bar is 20 nm.  (c) STM image showing the atomic resolution of the hexagonal lattice of monolayer graphene (0.5 V, 100 pA, 10 nm $\times$ 10 nm). The scale bar is 2 nm. }
\end{figure}

Figure \ref{fig:schematic}(a) shows a schematic diagram of the measurement setup. The bias voltage is applied to the STM tip and the sample is grounded. The particle on the tip apex is a Au nanoparticle attached after the tip conditioning on the Au surface. Because of the cleanliness of the samples, we are able to obtain large area images of the graphene surface which are free of defects or residue from sample processing. Figure \ref{fig:schematic}(b) shows the optical microscope image of the monolayer graphene flake with gold contacts on the SiO$_2$ substrate. Figure \ref{fig:schematic}(c) shows a typical 100 $\times$ 100 nm area of the graphene surface.  The sample is smooth with a slowly varying height due to the underlying SiO$_2$ substrate.  There are no sharp features that arise from the presence of contaminants on the surface or defects.  Atomic resolution topography of the sample is shown in Fig. \ref{fig:schematic}(d) showing the hexagonal lattice of carbon atoms as expected for monolayer graphene along with height variation due to the uneven SiO$_2$ substrate \cite{ishigami2007,stolyarova2007,deshpande2009,zhang2008,zhang2009}.

\begin{figure}[t]
\includegraphics[width=0.5 \textwidth]{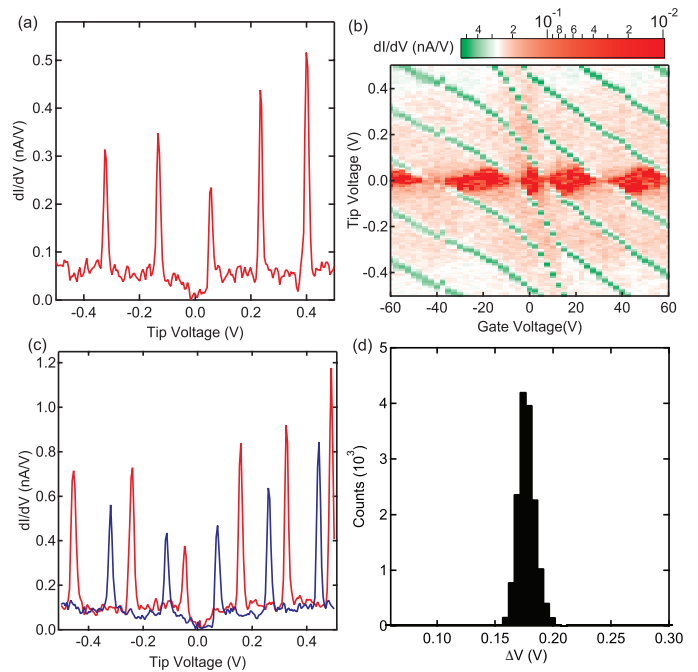}
\caption{\label{fig:CB} (color online) (a) dI/dV spectroscopy showing peaks due to the addition of individual electrons to the nanoparticle. (b) dI/dV as a function of tip voltage and gate voltage, showing a series of diamonds.  This demonstrates that the nanoparticle on the end of the tip acts as a quantum dot. (c) dI/dV curves taken at two different points on the graphene surface.  Each shows a series of peaks but they are horizontally offset due to the different potential on the graphene surface.  (d) Distribution of peak spacings for a 100 nm $\times$ 100 nm area of the graphene film.  The peaks are clustered around the same value showing that there is a single quantum dot located on the end of the tip.}
\end{figure}

We have measured the local density of states (LDOS) as a function of energy. For this measurement, the tip is held at a fixed location and the feedback loop is turned off. The voltage between the tip and substrate is ramped and the differential conductance as a function of voltage is acquired. An ac modulation voltage of 5 mV rms at 570 Hz is applied to the tip and the resulting current is detected with a lockin amplifier. Figure \ref{fig:CB} (a) shows a typical dI/dV curve acquired with a nanoparticle at the end of the tip.  There are a series of sharp peaks that are equally spaced in energy.  This is in contrast to a measurement on graphene without the nanoparticle which shows a smoothly varying density of states \cite{zhang2008, deshpande2009}.  The peaks arise due to the addition of individual electrons to the nanoparticle.  The setpoint current and voltage sets the overall resistance between the tip and graphene.  Using a small setpoint current ensures that the resistance between the nanoparticle and graphene, R$_{sub}$, is always much greater than the resistance between the nanoparticle and tip, R$_{tip}$.  Because of this strong asymmetry in the tunnel barriers, the barrier between the nanoparticle and graphene is always the rate limiting barrier.  Therefore, we obtain a Coulomb staircase with the spacing of the peaks determined by the capacitance between the nanoparticle and graphene \cite{Hanna}.

We have measured the dI/dV curves for different back gate voltages on the monolayer graphene sample. Once again the curves are characterized by sharp peaks which vary as a function of gate voltage. Figure \ref{fig:CB} (b) plots the dI/dV curves as a function of gate voltage and tip voltage.  It shows the characteristics of a Coulomb blockade diamond plot providing further evidence of the presence of a small nanoparticle on the tip.  In the region around zero sample voltage, the bright red diamond shaped regions show that there is no flow of current, indicating Coulomb blockade. As the voltage increases a series of peaks appear.  Between each peak the number of electrons on the nanoparticle changes by one. The gate acts to induce charge on the nanoparticle and therefore the peaks change in energy.  Only one set of peaks is strong, this is due to the asymmetry of the tunnel barriers. This strong peak corresponds to lining up the energy of a state on the nanoparticle with the energy of the graphene.  Therefore, as the voltage between the tip and graphene is increased, we see a series of peaks separated by $e/C_{sub}$ where $C_{sub}$ is the capacitance between the nanoparticle and graphene.  The strong peaks slope towards decreasing tip voltage as the gate voltage is increased.  This is the opposite direction expected for a normal gate electrode which would tend to induce negative charges as the gate voltage increases.  However, the effect of the gate on the nanoparticle is screened by the graphene.  A positive voltage on the gate induces electrons on the graphene which then tend to shift the energy levels on the nanoparticle.  Effectively, the graphene layer acts to reverse the sign of the gate voltage.

As the tip is moved to different positions of the graphene, the energies where peaks occur shift.  This is seen in Fig. \ref{fig:CB} (c) which shows the spectroscopy results from two locations on the graphene surface separated by a distance of 10 nm.  The entire dI/dV curve has been shifted along the tip voltage axis. All of the peaks still have the same spacing showing that the capacitance between the nanoparticle and graphene remains unchanged.  However, the energy at which the peaks occur has changed.  This is due to the presence of an offset charge on the nanoparticle.  Below we will show that this offset arises due to variations in the potential on the surface of the graphene.  Figure \ref{fig:CB} (d) shows the distribution of peak spacings over a 100 nm x 100 nm area.  As the peak spacings cluster around a single value, this shows that the we are using the same nanoparticle for all of our measurements.  On the other hand, the distribution of the energies of the peaks over the same area shows a uniform distribution.  When a new nanoparticle is attached to the tip, the peak spacing changes but otherwise the results are unchanged.

\begin{figure}[t]
\includegraphics[width=0.5 \textwidth]{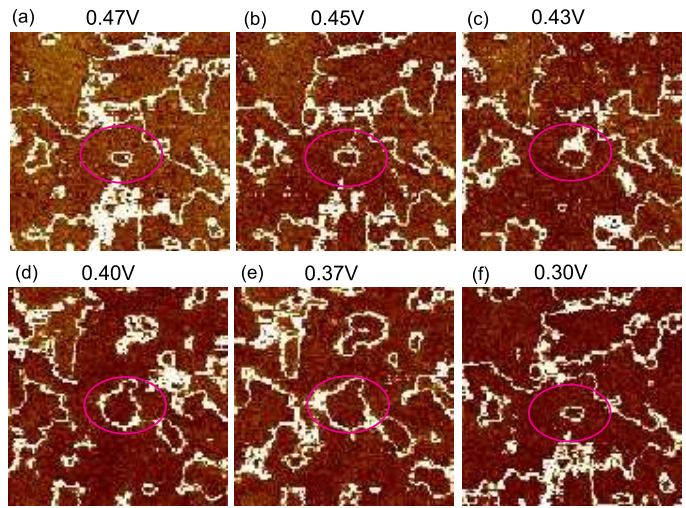}
\caption{\label{fig:2DMap} (color online) Local density of states maps taken at a series of tip voltages as indicated for each map. Each map is 100 nm $\times$ 100 nm. A ring is circled to show how it changes with energy. They start out circular (a-b) and slowly become asymmetric (c-e). Then a second ring forms (f) at the same location once the charge on the nanoparticle changes by one.}
\end{figure}

Figure \ref{fig:2DMap} shows the local density of states over a 100 nm by 100  nm area of the sample.  Six images from a movie of the density of states as a function of tip voltage are shown\cite{movie}.  The bright regions represent the locations where electrons are added to the nanoparticle.  They are peaks in the dI/dV curve.  Following one of the bright lines corresponds to moving along a path of constant induced charge on the nanoparticle.  As the voltage between the tip and graphene is changed, the location where electrons are added to the nanoparticle also changes.  Figure \ref{fig:2DMap} (a) to (f) shows the evolution of the rings as the tip voltage is changed.  Notice that the circled ring grows as the voltage goes from 0.47 V to 0.37.  When the sample voltage has reached 0.30 V, Fig.  \ref{fig:2DMap} (f), an additional electron has been added to the nanoparticle and the image is equivalent to Fig. \ref{fig:2DMap} (a) at 0.47 V.  The spatial variation in the density of states map implies that the electrostatic environment of the nanoparticle is changing as a function of position.  Therefore, these maps give information about the local potential or charge density fluctuations in graphene as discussed below.

\begin{figure}[t]
\includegraphics[width=0.35 \textwidth]{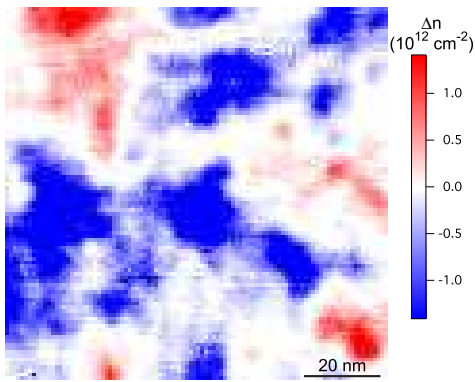}
\caption{\label{fig:ChargeMap} (color online) Charge density fluctuations in the same 100 nm by 100 nm region of the flake as in Fig. \ref{fig:2DMap}.  Red areas are positively charged while blue areas are negatively charged.}
\end{figure}

To obtain a map of the local potential or equivalently the charge density, we can analyze the energy at which the peaks occur as a function of position.  Since the peaks are arising due to Coulomb blockade, we expect that the spacing in energy between peaks will always remain constant at $e/C_{sub}$.  This is the reason that Fig. \ref{fig:2DMap} (a) and (f) look identical; they are separated in energy by $e/C_{sub}$.  On the other hand, the energy at which the peaks occur gives information about the amount of charge which must be induced on the nanoparticle in order for it to conduct.  If the offset charge on the nanoparticle is $e/2$, then we expect a series of peaks at $ne/C_{sub}$ where n is the number of electrons induced on the nanoparticle.  As the offset charge changes, the energies of the peaks will shift such that they occur at $(\Delta Q + ne)/C_{sub}$ where $\Delta Q$ is the offset charge relative to $e/2$.  In both cases, the peaks are spaced by $e/C_{sub}$.  We have used the data in Fig. \ref{fig:2DMap} to measure the offset charge at each point on the graphene surface.    This offset charge can be converted to a potential difference, $\Delta V$, by dividing by $C_{sub}$.  Furthermore, the potential difference can be converted to a local electron density using the band structure of graphene.  Using the linear dispersion relation of graphene, we find that fluctuations in charge density are related to fluctuations in potential by $\Delta n = (\Delta V)^2/\pi(\hbar v_F)^2$ where $v_F = 1.1 \times 10^6$ m/s is the Fermi velocity.  The results are shown in Fig. \ref{fig:ChargeMap}.  There are regions of positively and negatively charged puddles on the surface of the graphene.  This map shows a typical size scale of electron and hole puddles of 20 nm.  The regions of positive charge correspond to areas of the graphene flake where the rings in Fig. \ref{fig:2DMap} are shrinking as the tip voltage is decreased.  Likewise, rings that grow with decreasing tip voltage, such as the ring in the pink ellipse in Fig. \ref{fig:2DMap} are areas of negative charge on the graphene.

We find that the FWHM of our distribution of charge densities is $\Delta n = 3 \times 10^{11}$ cm$^{-2}$.  This is about a factor of 6 higher than found in previous SET measurements of electron and hole puddles on graphene\cite{martin}.  However, our significantly improved spatial resolution leads to the higher variation because we are now sensitive to charge fluctuations on the nanometer scale as opposed to the previous measurements which were limited to about 150 nm resolution.  With the lower spatial resolution, the charge fluctuations tend to be averaged away.  Our value for the charge fluctuation is similar to previous STM measurements of electron and hole puddles\cite{zhang2009}, however this technique has the ability to resolve much smaller changes in charge density because of the high energy sensitivity of the Coulomb blockade spectroscopy.

\begin{figure}[b]
\includegraphics[width=0.45 \textwidth]{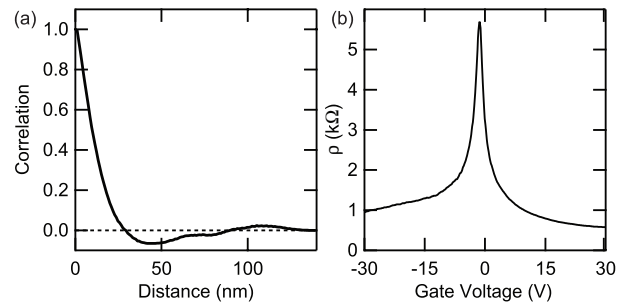}
\caption{\label{fig:Theory} (a) Correlation length for the puddles shown in Fig. \ref{fig:ChargeMap}.(b) Resistivity as a function of gate voltage for the graphene flake.}
\end{figure}

By performing an auto-correlation of the image in Fig. \ref{fig:ChargeMap}, we can find the correlation length of the puddles.  The results are shown in Fig. \ref{fig:Theory} (a).  It shows a FWHM of the distribution of 20 nm.  Theoretical calculations have shown that scattering from random charged impurities causes electron and hole puddles and their correlation length varies with the concentration of these impurities \cite{rossi2008}.  We can estimate the number of impurities in our sample from the electrical conductivity at high gate voltages \cite{adam2007}.  The conductivity is given by $\sigma=20 e \epsilon V_g / (h n_i t)$ where $n_i$ is the impurity density, $\epsilon$ is the dielectric constant of SiO$_2$, $t$ is the oxide thickness and $V_g$ is the gate voltage.  Figure \ref{fig:Theory} (b) shows the resistivity as a function of gate voltage.  From this plot, we can estimate the density of impurities as $n_i \approx 1\times 10^{11}$cm$^{-2}$.  This impurity concentration gives a theoretical correlation length for the puddles of 12 nm.  The global nature of the transport measurement compared to our local measure of the charge density fluctuations may explain the small discrepancy between the theoretical puddle size and our images.

Thus we have shown that a nanoparticle at the STM tip apex can act as a sensitive charge sensor to detect the charge density fluctuations in graphene. Using spatially resolved spectroscopy measurements we have mapped these fluctuations and found a characteristic puddle size of 20 nm.  A comparison with theoretical calculations shows that our graphene samples have an impurity density on the order of $10^{11}$ cm$^{-2}$ which give rise to these puddles.  In order for graphene devices to reach their full potential the number of impurities must be significantly reduced.  This technique of using a gold nanoparticle at the tip apex for spatially resolved measurements will enable us to detect the reduction in the number of impurities and consequently test the quality of the flakes for device fabrication.

We are thankful for discussions with Philippe Jacquod. AD and BJL acknowledge the support of the U. S. Army Research Laboratory and the U. S. Army Research Office under contract/grant number W911NF-09-1-0333 and NSF EECS/0925152.  CNL, WB and FM acknowledge the support of NSF CAREER DMR/0748910, NSF CBET/0756359 and ONR/DMEA Award H94003-07-2-0703.

\end{document}